# A Double-modulation Effect Detected in a Double-mode High-amplitude $\delta$ Scuti Star: KIC 10284901

Tao-Zhi Yang[1,2] and Ali Esamdin[1,2,3]
[1] Xinjiang Astronomical Observatory, Chinese Academy of Science, Urumqi 830011, People's Republic of China; aliyi@xao.ac.cn
[2] University of Chinese Academy of Sciences, Beijing 100049, People's Republic of China; yangtaozhi2018@163.com

## Abstract

In this paper, we present an analysis of the pulsating behavior of *Kepler* target KIC 10284901. The Fourier transform of the high-precision light curve reveals seven independent frequencies for its light variations. Among them, the first two frequencies are main pulsation modes: F0 = 18.994054(1) day$^{-1}$ and F1 = 24.335804(4) day$^{-1}$; the ratio F0/F1 = 0.7805 classifies this star as a double-mode high-amplitude $\delta$ Scuti (HADS) star; another two frequencies, $f_{m1}$ = 0.4407 day$^{-1}$ and $f_{m2}$ = 0.8125 day$^{-1}$, are detected directly, and the modulations of $f_{m1}$ and $f_{m2}$ to F0 and F1 modes (seen as quintuplet structures centered on these two modes in the frequency spectrum) are also found. This is the first detection of a double-modulation effect in the HADS stars. The features of the frequency pattern and the ratio ($f_{m1}/f_{m2} \approx$ 1:2), as well as the cyclic variation of amplitude of the two dominant pulsation modes, which seem to be similar to that in Blazhko RR Lyrae stars, indicate this modulation might be related to the Blazhko effect. A preliminary analysis suggests that KIC 10284901 is in the bottom of the HADS instability strip and situated in the main sequence.

*Key words:* stars: oscillations – stars: variables: delta Scuti

## 1. Introduction

The *Kepler Space Telescope* is designed to search for terrestrial planets orbiting the solar-type stars by the transits method (Borucki et al. 2010; Koch et al. 2010). As a supporting program, the *Kepler* asteroseismolog program possesses an intrinsic important role in the core planet search project (Gilliland et al. 2010). Owing to the ultra-high photometric precision data at the level of $\mu$mag, the *Kepler* mission has significantly improved our understanding of different types of variable stars (e.g., Bedding et al. 2011; Giammichele et al. 2018); and the continuous observations spanning about 4 yr provide an excellent opportunity to monitor the amplitude modulation of different pulsators (e.g., Benkó et al. 2014; Bowman & Kurtz 2014; Bowman et al. 2016). At present, the *Kepler* mission has found at least 2000 $\delta$ Sct stars (Balona & Dziembowski 2011; Balona 2014; Bowman et al. 2016). Among them, some stars show amplitude modulation of pulsation modes caused for different reasons, e.g., beating, mode coupling, and rotation (e.g., Bowman & Kurtz 2014; Bowman et al. 2016; Yang et al. 2018b). These targets are excellent for asteroseismic study, as they could improve our knowledge of the stellar structure and evolution for stars.

High-amplitude $\delta$ Sct (HADS) stars are usually considered a subclass of $\delta$ Sct stars, with amplitudes of peak-to-peak light variations larger than 0.3 mag (Breger 2000). From the relationship between the amplitude and the measured rotational velocity ($v \sin i$) provided by Breger (2000), HADS stars are typical slow rotators with $v \sin i \leqslant 30$ km s$^{-1}$. They seem to concentrate in the center part of the $\delta$ Sct instability region, and occupy a relatively narrow strip with a width of 300 K in temperature (McNamara 2000). As seen with ground-based observations, HADS stars are usually pulsating with only one or two radial modes (e.g., YZ Boo: Yang et al. 2018a; KIC 5950759: Yang et al. 2018b, etc). In recent decades, some stars also been found to exhibit nonradial modes with low amplitude, owing to extensive photometric campaigns. With the advent of space missions, especially the *Kepler* mission, more interesting phenomena have been discovered, including low-amplitude pulsation modes and long-term variations in pulsating stars (Balona et al. 2012; Bowman & Kurtz 2014). Several stars observed by *Kepler* show triplet or quintuplet structures in their frequency spectra (Kolenberg et al. 2011; Benkó et al. 2014). For instance, in HADS star KIC 5950759, a pair of triplet structures centered on the main frequency were detected in its frequency spectra, and the cause of the triplet structure is inferred to be the amplitude modulation of stellar rotation, 0.3193 day$^{-1}$ ($v \sin i \approx 33$ km s$^{-1}$) (Yang et al. 2018b). These low-amplitude multiplet structures might improve our knowledge of the HADS stars and offer new clues for probing the stellar interior and physical processes.

KIC 10284901 ($\alpha_{2000}$ = 19$^h$43$^m$46$^s$.4, $\delta_{2000}$ = +47°20′32″.8, 2MASS: J19434637+4720323) was found to be a $\delta$ Scuti star in the RApid Temporal Survey of the *Kepler* field (RATS-*Kepler*) by Ramsay et al. (2014). In that survey, Ramsay et al. (2014) reported KIC 10284901 was a mid-late A type star and it might be a HADS star due to its high-amplitude light variations. Some basic properties of this star from that survey and the *Kepler* Input Catalog (KIC; Brown et al. 2011) are listed in Table 1. This star was also selected as a target in the *Kepler* Guest Observer program and continuously observed for more than 10 months in both long cadence (LC) with 29.4 minute effective integrations and short cadence (SC) with 58.8 s effective integrations (Gilliland et al. 2010). Due to the strong effect of signal averaging in LC data, we only use the SC data in this work. The unique and high-precision photometric data make KIC 10284901 an ideal source to deeply investigate its pulsation behavior.

[3] Corresponding author.

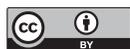

**Table 1**
Basic Properties of KIC 10284901

| Parameters | KIC 10284801 | References |
|---|---|---|
| $K_P$ | 15.459 | a |
| $P$ | 75.8 minutes | b |
| $G_{Gaia}$ | 15.433 | d |
| $g_{SDSS}$ | 15.507 | a |
| $i_{SDSS}$ | 15.468 | a |
| $z_{SDSS}$ | 15.543 | a |
| D51 | 15.535 | a |
| $J_{2MASS}$ | 14.810 | a |
| $H_{2MASS}$ | 14.611 | a |
| $K_{2MASS}$ | 14.525 | a |
| g | 15.730 | c |
| $U - g$ | 0.06 | c |
| $g - r$ | 0.32 | c |
| $E(G_{BP} - G_{RP})$ | 0.46 | d |
| $A_G$ | 0.21 | d |
| $T_{KIC}$ | $8420 \pm 250$ K | a |
| $T_{GTC}$ | $7710 \pm 180$ K | b |
| $T_{Gaia}$ | $7483 \pm 324$ K | d |
| log g | $4.0 \pm 0.25$ dex | a |

**Note.** (a) KIC Brown et al. (2011). (b) Ramsay et al. (2014). (c) Greiss et al. (2012). (d) These parameters are available in the *Gaia* Archive (http://gea.esac.esa.int/archive/), and the uncertainty of the effective temperature is a typical value (Andrae et al. 2018).

## 2. Observations and Data Reduction

KIC 10284901 was observed continuously from BJD 2456107.13 to 2456390.97 (Q14.1: 34,809 SC observations; Q14.3: 49,957 SC observations; Q16.1: 7622 SC observations;, Q16.3: 46,461 SC observations; 138,849 SC observations in total) spanning 283.84 days. The Kepler Asteroseismic Science Operations Center (KASOC) database[4] (Kjeldsen et al. 2010) provides the SC photometric flux data of KIC 10284901 as two types: one is the "raw" data, which in fact has been reduced by the NASA *Kepler* Science pipeline, and the other is the flux data corrected by KASOC Working Group 4 (WG#4: $\delta$ Scuti targets). We use the corrected flux and perform corrections, including eliminating outliers, as well as the possible linear trends in some quarters. The flux data are converted to the magnitude scale, then the mean value of each quarter is subtracted, and the rectified time series is obtained. A portion of the rectified SC light curve is shown in Figure 1. It is clear that the light amplitude of KIC 10284901 is larger than 0.3 mag in SC data. Inspection of the light curve indicates that the amplitude has a repeat of about 2.5 days.

## 3. Frequency Analysis

We performed a Fourier transform for the rectified SC light curve using PERIOD04 (Lenz & Breger 2005). The light curve is fitted with the following formula:

$$m = m_0 + \Sigma A_i \sin(2\pi(f_i t + \phi_i)), \quad (1)$$

where $m_0$ denotes the zero-point, $A_i$ is the amplitude, $f_i$ is the frequency, and $\phi_i$ is the corresponding phase.

To detect more significant frequencies, a frequency range of $0 < \nu < 50 \, \text{day}^{-1}$, which covers the typical pulsation frequency of the $\delta$ Sct stars, was chosen in this work. In the extraction of significant frequencies, the highest peak in the frequency spectrum was considered as a significant frequency, then a multi-frequency least-squares fit using formula 1 was conducted for the light curve with all the significant frequencies detected, resulting in the solutions of all the significant frequencies. Next, all the frequencies of combination signals are fixed to the exact values they are supposed to be, and only leave the independent frequencies, all amplitudes, and all phases as free parameters to be improved. A constructed light curve using the above solutions was subtracted from the data, and the residual was obtained to search for a significant term in next step. The above steps were repeated until there was no significant peak in the residual. The criterion (S/N > 4.0) suggested by Breger et al. (1993) was adopted to judge the significant peaks. The uncertainties of frequencies were calculated following Montgomery & Odonoghue (1999).

---

[4] KASOC database: http://kasoc.phys.au.dk.

A total of 151 significant frequencies were extracted in this work, and a full list of the extracted frequencies ($f_{S1}$ to $f_{S151}$), with their corresponding amplitude and identifications, is given in Table 2. After pre-whitening of the 151 frequencies, the amplitude spectrum of the residual is shown in Figure 2. No peak is statistically significant in the residual and the overall distribution of the residual has typical noise.

Seven high-amplitude independent frequencies were detected, five of which were found in the range of 18–25 day$^{-1}$, and the other two (i.e., $f_{S3}$ and $f_{S4}$) were in the range of 0–1 day$^{-1}$. Among these independent frequencies, $f_{S1}$ and $f_{S2}$ give a ratio of 0.7805, which is in the typical period ratio range of the first overtone and the fundamental mode for the double-mode HADS star. If the highest frequency $f_{S1}$ is assumed to be the fundamental mode, then KIC 10284901 is classified as a double-mode HADS star. We thus marked the frequencies $f_{S1}$ and $f_{S2}$ with "F0" and "F1" in the last column of Table 2, respectively. Petersen & Christensen-Dalsgaard (1996) presented a detailed diagram of double-mode HADS stars of different metallicities in their Figure 3. Their study showed that higher values of the P(F1)/P(F0) ratio are found for metal-poor stars. For KIC 10284901, the higher period ratio (0.7805) indicates that it may belong to Population II stars.

$f_{S3}$ and $f_{S4}$ are interesting, as they are out of the typical frequency range of the $\delta$ Scuti stars, and they are not combinations of F0 and F1. Considering a repeating with about 2.5 days in the light curve, the frequency $f_{S3}$ was marked with $f_{m1}$, and $f_{S4}$ with $f_{m2}$ in Table 2, respectively. Three other frequencies, $f_{S5}$, $f_{S6}$, and $f_{S7}$, were also considered as independent frequencies, as they are neither any combinations nor harmonics of other frequencies. Thus, we marked these three frequencies with "independent" in the last column of Table 2. Another eight frequencies (marked with "T1" and "T2" in the last column of Table 2) can be divided into two groups: one group consists of the frequencies marked with "T1," which are combinations of $f_{m1}$ and the main frequencies (i.e., F0 and F1); the other group is composed of the frequencies marked with "T2," which are combinations of $f_{m2}$ and the main frequencies.

## 4. Discussion

The four pairs of side peaks around the main frequencies F0 and F1 (i.e., (1) $f_{S8} = 19.43478 \, \text{day}^{-1}$ and $f_{S9} = 18.55332 \, \text{day}^{-1}$, centered on F0 with an interval of $f_{m1}$, (2) $f_{S10} = 19.80665 \, \text{day}^{-1}$ and $f_{S11} = 18.18152 \, \text{day}^{-1}$, centered on F0 with an interval of $f_{m2}$ shown in the left panel of Figure 3; (3) $f_{S12} = 24.7766 \, \text{day}^{-1}$ and $f_{S13} = 23.8952$ day$^{-1}$, centered on F1 with an interval of $f_{m1}$; and (4) $f_{S14} = 25.1483$ day$^{-1}$ and $f_{S15} = 23.5233$ day$^{-1}$, centered on F1 with an interval of $f_{m2}$, shown in the right panel of Figure 3), are the most interesting features in the frequency spectrum of KIC 10284901. Also listed in Table 2 are some other visible peaks that are not labeled in Figure 3, $f_{S5}$ and $f_{S67}$, which are the peaks to the left of $f_{S8}$, and the three peaks $f_{S6}$, $f_{S37}$, and $f_{S123}$, which are to the left of $f_{S12}$.

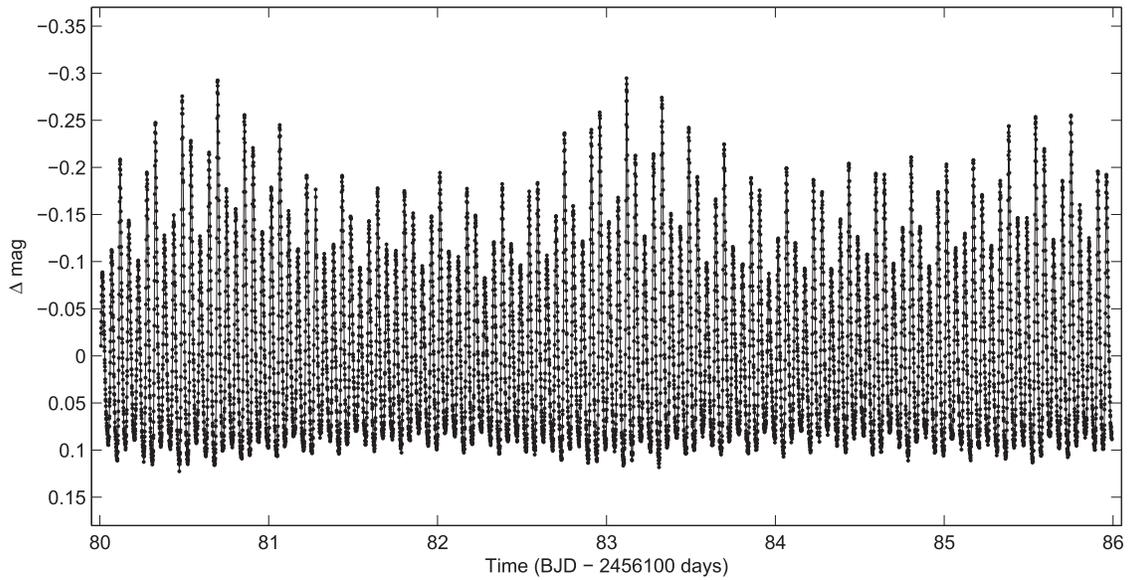

**Figure 1.** Portion of the SC light curve of KIC 10284901. Inspection of the light curve indicates that the amplitude has a repeat of about 2.5 days.

The side peaks around F0 and F1 in the frequency spectra of KIC 10284901 form four pairs of uniformly spaced triplets with intervals of $f_{m1} = 0.4407$ day$^{-1}$ and $f_{m2} = 0.8125$ day$^{-1}$. To investigate these side peaks, we first checked the possibility that these triplets were from the instrumental effects of *Kepler*, as performed by Yang et al. (2018b), and found that none of the known frequencies of instrumental effects from *Kepler* were equal to $f_{m1}$ or $f_{m2}$. Hence, the side peaks detected in the frequency spectra of this star are not caused by the known instrumental effects.

### 4.1. The Modulations of F0 and F1

*Kepler* has found that the multiplet structures can be shown in the frequency spectra of different types of pulsating variables (e.g., Kolenberg et al. 2011; Benkó et al. 2014; Yang et al. 2018b). In examining the δ Sct stars observed by *Kepler*, Breger et al. (2011) noted that in several stars the equally spaced frequency components were presented in their frequency spectra and these multiplet structures were considered to be from the stellar intrinsic variations. With asteroseismology, these equally spaced frequencies might provide more information about the global properties. Hence, it is astrophysically interesting to explore the nature of the triplet structures in the frequency spectra of KIC 10284901.

#### 4.1.1. New Radial Modes or Nonradial Modes?

For the equidistant or nearly equidistant frequency triplets shown in the spectra of the pulsating stars, Breger & Kolenberg (2006) proposed an explanation called "The Combination Mode Hypothesis." In this scenario, the frequency $f_{S8}$ (=19.43478 day$^{-1}$) detected in the SC spectrum of KIC 10284901 can be regarded as the third mode (F0 and F1 are the fundamental and the first overtone mode, respectively), and $f_{S9}$ (=18.55332 day$^{-1}$) can be considered the combination of 2F0 and $f_{S8}$, i.e., $f_{S9} = 2F0 - f_{S8}$. Some other frequencies are therefore the combinations of F0, F1, and $f_{S8}$. Similarly, $f_{S11}$ (=18.18152 day$^{-1}$) can be regard as the fourth mode and some frequencies are also combinations of F0, F1, and $f_{S11}$. Thus, KIC 10284901 might be a multimode radial variable. However, the ratios of F0/$f_{S8}$ (=0.9773) and F0/$f_{S11}$ (=1.0447) are far away from the typical ratio for the $P_{2O}/P_F$ (0.611 ∼ 0.632) and $P_{3O}/P_F$ (0.500 ∼ 0.525)($P_F$ is the period of the fundamental radial mode, and $P_{2O}$ and $P_{3O}$ are the periods of the second and third radial overtones, respectively) (Stellingwerf 1979). This seems to rule out the possibility that these two frequencies belong to the radial modes.

If $f_{S8}$ or $f_{S11}$ is assumed to be a nonradial mode, the equidistant triplet structure formed by the nonradial modes can occur only when the stars rotate with an extremely low velocity. When the stars rotate only about a few km s$^{-1}$, the rotationally split multiplet will not be symmetric, and the equidistant structure in the spectrum will also not be seen (Pamyatnykh 2000). Consequently, $f_{S8}$ and $f_{S11}$ are unlikely to be nonradial modes of KIC 10284901. The side peaks might be caused by some modulation effects.

#### 4.1.2. Amplitude Modulation with Rotation?

Rotation plays an important role in stellar evolution and pulsations (Pamyatnykh 2000). Several δ Scuti stars found in the *Kepler* mission show equidistant multiplets in their frequency spectra that were caused by modulation of the amplitude with stellar rotation (Breger et al. 2011). A low-amplitude modulation frequency of 0.16 day$^{-1}$ to the dominant frequencies was detected in KIC 9700322 and it was confirmed to be the stellar rotation frequency by the high-dispersion spectral observations (Breger et al. 2011).

Another example is KIC 11754974 (Murphy et al. 2013), which is a metal-poor double-mode HADS star. The light curve of this star shows apparent amplitude modulation, similar to that in KIC 10284901. The Fourier transform of its light curve also reveals a high number of combination frequencies of the

Table 2
All Frequencies Detected in SC Data (Denoted by $f_{Si}$)

| $f_{Si}$ | Frequency (day$^{-1}$) | Amplitude (mmag) | S/N | Identification |
|---|---|---|---|---|
| | Independent frequencies | | | |
| 1 | 18.994054 ± 0.000001 | 115.038 | 344.4 | F0 |
| 2 | 24.335804 ± 0.000004 | 22.463 | 105.6 | F1 |
| 3 | 0.44069 ± 0.00002 | 3.979 | 17.2 | $f_{m1}$ |
| 4 | 0.81249 ± 0.00002 | 3.627 | 38.2 | $f_{m2}$ |
| 5 | 19.36979 ± 0.00003 | 3.273 | 22.2 | Independent |
| 6 | 24.59564 ± 0.00003 | 3.101 | 24.8 | Independent |
| 7 | 23.17789 ± 0.00004 | 2.015 | 36.9 | Independent |
| | Equally spaced frequencies | | | |
| 8 | 19.434777 ± 0.000008 | 10.855 | 40.6 | F0+$f_{m1}$, "T1" |
| 9 | 18.55332 ± 0.00005 | 1.664 | 20.8 | F0−$f_{m1}$, "T1" |
| 10 | 19.80665 ± 0.00008 | 1.086 | 18.6 | F0+$f_{m2}$, "T2" |
| 11 | 18.18152 ± 0.00002 | 4.560 | 34.3 | F0−$f_{m2}$, "T2" |
| 12 | 24.7766 ± 0.0002 | 0.340 | 8.5 | F1+$f_{m1}$, "T1" |
| 13 | 23.8952 ± 0.0001 | 0.715 | 14.3 | F1−$f_{m1}$, "T1" |
| 14 | 25.1483 ± 0.0002 | 0.372 | 8.4 | F1+$f_{m2}$, "T2" |
| 15 | 23.5233 ± 0.0002 | 0.475 | 11.5 | F1−$f_{m2}$, "T2" |
| | Combination frequencies | | | |
| 16 | 37.988110 ± 0.000002 | 32.759 | 213.6 | 2F0 |
| 17 | 5.341754 ± 0.000006 | 13.678 | 101.1 | F1−F0 |
| 18 | 43.329943 ± 0.000007 | 11.613 | 79.9 | F0+F1 |
| 19 | 43.58978 ± 0.00004 | 1.976 | 17.8 | F0+$f_{S6}$ |
| 20 | 48.67163 ± 0.00004 | 1.887 | 45.8 | 2F1 |
| 21 | 5.60170 ± 0.00005 | 1.539 | 20.5 | $f_{S6}$−F0 |
| 22 | 38.36385 ± 0.00006 | 1.462 | 16.7 | F0+$f_{S5}$ |
| 23 | 0.37577 ± 0.00006 | 1.286 | 12.6 | $f_{S5}$−F0 |
| 24 | 43.70589 ± 0.00008 | 1.055 | 24.9 | $f_{S5}$+F1 |
| 25 | 38.73945 ± 0.00009 | 0.878 | 27.3 | 2$f_{S5}$ |
| 26 | 42.1720 ± 0.0001 | 0.762 | 22.0 | F0+$f_{S7}$ |
| 27 | 4.1839 ± 0.0001 | 0.720 | 17.5 | $f_{S7}$−F0 |
| 28 | 48.9316 ± 0.0002 | 0.470 | 15.1 | F1+$f_{S6}$ |
| 29 | 4.9661 ± 0.0002 | 0.387 | 7.9 | F1−$f_{S5}$ |
| 30 | 0.2595 ± 0.0003 | 0.237 | 4.3 | $f_{S6}$−F1 |
| 31 | 47.5139 ± 0.0005 | 0.173 | 7.5 | F1+$f_{S7}$ |
| 32 | 22.7373 ± 0.0006 | 0.124 | 5.0 | $f_{S7}$−$f_{m1}$ |
| 33 | 13.65236 ± 0.00001 | 7.256 | 85.3 | 2F0−F1 |
| 34 | 38.42881 ± 0.00002 | 4.901 | 36.0 | 2F0+$f_{m1}$ |
| 35 | 37.17565 ± 0.00004 | 2.328 | 42.5 | 2F0−$f_{m2}$ |
| 36 | 29.67758 ± 0.00005 | 1.509 | 51.1 | 2F1−F0 |
| 37 | 24.71221 ± 0.00006 | 1.385 | 19.3 | F1−F0+$f_{S5}$ |
| 38 | 4.90113 ± 0.00006 | 1.267 | 23.1 | F1−F0−$f_{m1}$ |
| 39 | 43.77050 ± 0.00007 | 1.157 | 25.6 | F0+F1+$f_{m1}$ |
| 40 | 13.39238 ± 0.00009 | 0.934 | 22.4 | 2F0−$f_{S6}$ |
| 41 | 38.29677 ± 0.00009 | 0.869 | 27.8 | 2$f_{S5}$−$f_{m1}$ |
| 42 | 5.7180 ± 0.0001 | 0.771 | 16.8 | $f_{S5}$+F1−2F0 |
| 43 | 19.7454 ± 0.0001 | 0.678 | 17.5 | 2$f_{S5}$−F0 |
| 44 | 18.6183 ± 0.0001 | 0.623 | 10.6 | 2F0−$f_{S5}$ |
| 45 | 6.1543 ± 0.0001 | 0.578 | 14.9 | F1−F0+$f_{m2}$ |
| 46 | 42.8892 ± 0.0000 | 0.426 | 11.6 | F0+F1−$f_{m1}$ |
| 47 | 14.8101 ± 0.0002 | 0.428 | 14.6 | 2F0−$f_{S7}$ |
| 48 | 37.5475 ± 0.0002 | 0.369 | 8.7 | 2F0−$f_{m1}$ |
| 49 | 5.7824 ± 0.0002 | 0.356 | 9.8 | F1−F0+$f_{m1}$ |
| 50 | 42.5175 ± 0.0002 | 0.344 | 11.7 | F0+F1−$f_{m2}$ |
| 51 | 4.5292 ± 0.0003 | 0.324 | 9.8 | F1−F0−$f_{m2}$ |
| 52 | 38.8007 ± 0.0003 | 0.300 | 11.9 | 2F0+$f_{m2}$ |
| 53 | 14.0281 ± 0.0003 | 0.298 | 8.7 | $f_{S5}$−F1+F0 |
| 54 | 0.7514 ± 0.0003 | 0.285 | 4.3 | 2$f_{S5}$−2F0 |
| 55 | 14.4037 ± 0.0003 | 0.244 | 10.5 | 2$f_{S5}$−F1 |
| 56 | 3.7432 ± 0.0003 | 0.239 | 8.8 | $f_{S7}$−F0−$f_{m1}$ |
| 57 | 18.7342 ± 0.0003 | 0.243 | 6.3 | F0+F1−$f_{S6}$ |
| 58 | 23.9601 ± 0.0003 | 0.237 | 6.6 | F0+F1−$f_{S5}$ |
| 59 | 17.8352 ± 0.0004 | 0.222 | 4.8 | $f_{S7}$−F1+F0 |
| 60 | 19.2539 ± 0.0004 | 0.219 | 7.2 | F0+$f_{S6}$−F1 |
| 61 | 44.0309 ± 0.0005 | 0.174 | 5.6 | F0+$f_{S6}$+$f_{m1}$ |

Table 2
(Continued)

| $f_{Si}$ | Frequency (day$^{-1}$) | Amplitude (mmag) | S/N | Identification |
|---|---|---|---|---|
| 62 | 20.2476 ± 0.0004 | 0.185 | 4.0 | F0+$f_{m1}$+$f_{m2}$ |
| 63 | 29.9374 ± 0.0004 | 0.188 | 6.1 | F1−F0+$f_{S6}$ |
| 64 | 44.1422 ± 0.0006 | 0.130 | 4.1 | F0+F1+$f_{m2}$ |
| 65 | 4.9964 ± 0.0006 | 0.144 | 4.4 | $f_{S7}$−F0+$f_{m2}$ |
| 66 | 32.64640 ± 0.00004 | 2.323 | 53.9 | 3F0−F1 |
| 67 | 19.30291 ± 0.00004 | 1.919 | 32.1 | 2$f_{S5}$−F0−$f_{m1}$ |
| 68 | 14.09292 ± 0.00008 | 1.055 | 26.4 | 2F0−F1+$f_{m1}$ |
| 69 | 10.6835 ± 0.0001 | 0.819 | 36.9 | 2F1−2F0 |
| 70 | 12.8399 ± 0.0002 | 0.446 | 13.7 | 2F0−F1−$f_{m2}$ |
| 71 | 32.3866 ± 0.0003 | 0.297 | 11.3 | 3F0−$f_{S6}$ |
| 72 | 33.3978 ± 0.0004 | 0.206 | 7.4 | F0+2$f_{S5}$−F1 |
| 73 | 4.5903 ± 0.0004 | 0.200 | 7.6 | F0−2$f_{S5}$+F1 |
| 74 | 15.2508 ± 0.0004 | 0.196 | 10.7 | 2F0−$f_{S7}$+$f_{m1}$ |
| 75 | 37.6118 ± 0.0005 | 0.165 | 5.0 | 3F0−$f_{S5}$ |
| 76 | 24.2717 ± 0.0004 | 0.184 | 5.1 | F1−F0+$f_{S5}$−$f_{m1}$ |
| 77 | 13.9609 ± 0.0005 | 0.161 | 6.6 | 2$f_{S5}$−F1−$f_{m1}$ |
| 78 | 33.8049 ± 0.0005 | 0.163 | 7.3 | 3F0−$f_{S7}$ |
| 79 | 33.0222 ± 0.0005 | 0.172 | 6.5 | 2F0+$f_{S5}$−F1 |
| 80 | 29.2365 ± 0.0006 | 0.129 | 5.3 | 2F1−F0−$f_{m1}$ |
| 81 | 44.3413 ± 0.0006 | 0.124 | 4.5 | 2$f_{S5}$+$f_{S6}$−F0 |
| 82 | 14.4647 ± 0.0007 | 0.118 | 5.4 | 2F0−F1+$f_{m2}$ |
| 83 | 28.5876 ± 0.0004 | 0.204 | 7.5 | F1+$f_{S7}$+$f_{m1}$−$f_{S5}$ |
| 84 | 38.2478 ± 0.0006 | 0.142 | 5.2 | 2F0+$f_{S6}$−F1 |
| 85 | 0.3089 ± 0.0001 | 0.683 | 10.8 | 2$f_{S5}$−2F0−$f_{m1}$ |
| 86 | 8.3105 ± 0.0001 | 0.630 | 30.5 | 3F0−2F1 |
| 87 | 33.0871 ± 0.0002 | 0.485 | 16.4 | 3F0−F1+$f_{m1}$ |
| 88 | 13.2759 ± 0.0002 | 0.399 | 14.5 | 3F0−$f_{S5}$−F1 |
| 89 | 1.7424 ± 0.0002 | 0.433 | 6.0 | $f_{S7}$−F0−3$f_{m2}$ |
| 90 | 25.3469 ± 0.0003 | 0.301 | 9.5 | $f_{S6}$−2F0+2$f_{S5}$ |
| 91 | 44.7550 ± 0.0002 | 0.355 | 14.8 | 3$f_{S7}$−F1−$f_{m1}$ |
| 92 | 23.0762 ± 0.0003 | 0.294 | 8.4 | F0+F1−2$f_{m1}$−$f_{S5}$ |
| 93 | 31.8340 ± 0.0003 | 0.235 | 10.2 | 3F0−F1−$f_{m2}$ |
| 94 | 5.0330 ± 0.0004 | 0.208 | 6.0 | F1−2$f_{S5}$+F0+$f_{m1}$ |
| 95 | 5.2773 ± 0.0005 | 0.168 | 5.7 | $f_{S5}$+F1−2F0−$f_{m1}$ |
| 96 | 16.6799 ± 0.0004 | 0.181 | 6.8 | 2$f_{S7}$−2F1+F0 |
| 97 | 43.6392 ± 0.0004 | 0.190 | 6.0 | F1−F0+2$f_{S5}$−$f_{m1}$ |
| 98 | 18.2427 ± 0.0004 | 0.196 | 4.5 | 3F0−2$f_{S5}$ |
| 99 | 34.7890 ± 0.0006 | 0.142 | 6.0 | F1+2$f_{S6}$−2$f_{S5}$ |
| 100 | 25.7609 ± 0.0001 | 0.738 | 21.0 | 3$f_{S7}$−F0−F1−$f_{m1}$ |
| 101 | 18.6854 ± 0.0003 | 0.293 | 7.0 | 3F0−2$f_{S5}$+$f_{m1}$ |
| 102 | 27.3046 ± 0.0003 | 0.259 | 12.9 | 4F0−2F1 |
| 103 | 45.5959 ± 0.0005 | 0.161 | 6.3 | $f_{S6}$−F0+2$f_{S5}$+$f_{m1}$+$f_{m2}$ |
| 104 | 24.0270 ± 0.0005 | 0.156 | 5.5 | 2F0+F1−2$f_{S5}$+$f_{m1}$ |
| 105 | 35.6489 ± 0.0006 | 0.142 | 6.3 | 2F1+$f_{S6}$−2F0−$f_{m1}$+$f_{m2}$ |
| 106 | 13.7167 ± 0.0007 | 0.115 | 4.9 | 3F0+$f_{m1}$−$f_{S5}$−F1 |
| 107 | 32.2699 ± 0.0006 | 0.140 | 5.2 | 4F0−$f_{S5}$−F1 |
| 108 | 6.0934 ± 0.0006 | 0.125 | 4.6 | F1−3F0+2$f_{S5}$ |
| 109 | 17.8683 ± 0.0001 | 0.558 | 10.8 | 4F0−3$f_{S5}$ |
| 110 | 16.6548 ± 0.0002 | 0.362 | 12.2 | 2F1+$f_{S6}$−3F0−$f_{m1}$+$f_{m2}$ |
| 111 | 6.7667 ± 0.0003 | 0.302 | 13.9 | 3$f_{S7}$−2F0−F1−$f_{m1}$ |
| 112 | 9.3597 ± 0.0007 | 0.118 | 4.7 | 2F1−2F0−3$f_{m1}$ |
| 113 | 24.6049 ± 0.0002 | 0.341 | 9.4 | 4$f_{S7}$−F0−2F1−$f_{m1}$ |
| 114 | 12.2272 ± 0.0004 | 0.195 | 9.6 | 3F0−3$f_{S7}$+F1+$f_{m1}$ |
| 115 | 36.8624 ± 0.0004 | 0.179 | 6.2 | 5F0−3$f_{S5}$ |
| 116 | 23.1281 ± 0.0005 | 0.172 | 6.0 | 2F0+$f_{S6}$+$f_{S7}$+$f_{m1}$−2$f_{S5}$−F1 |
| 117 | 13.2396 ± 0.0006 | 0.143 | 5.0 | 2F1+$f_{S7}$−3F0−2$f_{m2}$ |
| 118 | 22.7643 ± 0.0006 | 0.138 | 4.9 | 3F0+F1−$f_{m1}$−3$f_{S5}$ |
| 119 | 23.0953 ± 0.0003 | 0.263 | 9.0 | $f_{S6}$+4F0−4$f_{S5}$ |
| 120 | 5.7546 ± 0.0003 | 0.231 | 6.6 | 4F0+2$f_{m2}$−2F1−$f_{S7}$ |
| 121 | 5.6111 ± 0.0005 | 0.173 | 5.6 | 4$f_{S7}$−2F0−2F1−$f_{m1}$ |
| 122 | 37.9959 ± 0.0004 | 0.178 | 5.6 | F0−$f_{S6}$+4$f_{S7}$−2F1−$f_{m1}$ |
| 123 | 24.7485 ± 0.0001 | 0.738 | 13.5 | 5F0+2$f_{m2}$−2F1−$f_{S7}$ |
| 124 | 18.8870 ± 0.0004 | 0.225 | 6.7 | 4$f_{S7}$+F0−3F1−$f_{m1}$−$f_{S5}$ |
| 125 | 28.7143 ± 0.0006 | 0.128 | 4.6 | 2F0+F1+3$f_{m1}$+$f_{S7}$−3$f_{S5}$ |

Table 2
(Continued)

| $f_{Si}$ | Frequency (day$^{-1}$) | Amplitude (mmag) | S/N | Identification |
| --- | --- | --- | --- | --- |
| 126 | 28.6591 ± 0.0002 | 0.422 | 13.8 | $f_{S6}+4F0-2F1-2f_{m1}+f_{m2}-f_{S7}$ |
| 127 | 43.7427 ± 0.0003 | 0.228 | 5.7 | $6F0+2f_{m2}-2F1-f_{S7}$ |
| 128 | 18.1570 ± 0.0001 | 0.706 | 9.3 | $6F0-3F1-f_{m1}+f_{m2}-f_{S7}$ |
| 129 | 47.6534 ± 0.0004 | 0.191 | 7.5 | $5F0+f_{S6}-2F1-2f_{m1}+f_{m2}-f_{S7}$ |
| 130 | 29.7254 ± 0.0004 | 0.184 | 7.9 | $3F1+3f_{S5}-4F0-f_{S6}-f_{m2}$ |
| 131 | 20.3316 ± 0.0004 | 0.183 | 4.3 | $4f_{S5}-3F0-3f_{m1}-f_{S16}+F1$ |
| 132 | 21.5457 ± 0.0005 | 0.176 | 6.4 | $3f_{S7}-2F1-f_{m1}-3F0+3f_{S5}$ |
| 133 | 2.2550 ± 0.0005 | 0.165 | 4.2 | $6f_{S5}-6F0$ |
| 134 | 37.2199 ± 0.0005 | 0.152 | 5.4 | $4F0-2F1+2f_{m1}+f_{S6}+f_{S7}-2f_{S5}$ |
| 135 | 1.3374 ± 0.0002 | 0.539 | 4.4 | $4f_{S5}-4F0-3f_{m1}-f_{S7}+F1$ |
| 136 | 37.1511 ± 0.0002 | 0.392 | 10.7 | $7F0-3F1-f_{m1}+f_{m2}-f_{S7}$ |
| 137 | 2.5516 ± 0.0004 | 0.215 | 5.2 | $3f_{S7}-2F1-f_{m1}-4F0+3f_{S5}$ |
| 138 | 20.0159 ± 0.0004 | 0.183 | 4.8 | $4F1-4F0-f_{S6}+2f_{m1}-f_{m2}+f_{S7}$ |
| 139 | 1.0216 ± 0.0002 | 0.374 | 4.2 | $4F1-5F0-f_{S6}+2f_{m1}-f_{m2}+f_{S7}$ |
| 140 | 17.6568 ± 0.0004 | 0.187 | 4.6 | $5F0-4f_{S5}+3f_{m1}+f_{S7}-F1$ |
| 141 | 24.6291 ± 0.0005 | 0.169 | 5.7 | $6F0+f_{S6}+2f_{m2}-3F1-f_{S7}-f_{S5}$ |
| 142 | 16.4421 ± 0.0006 | 0.128 | 5.3 | $5F0+2F1+f_{m1}-3f_{S7}-3f_{S5}$ |
| 143 | 17.9722 ± 0.0005 | 0.168 | 4.4 | $6F0-4F1+f_{S6}-2f_{m1}+f_{m2}-f_{S7}$ |
| 144 | 32.3149 ± 0.0006 | 0.138 | 5.6 | $5F0-3F1+3f_{m1}+f_{S6}+f_{S7}-2f_{S5}$ |
| 145 | 5.2586 ± 0.0006 | 0.124 | 4.4 | $6F0+f_{S6}+2f_{m2}-3F1-f_{S7}-2f_{S5}$ |
| 146 | 24.6680 ± 0.0002 | 0.414 | 7.6 | $2f_{S5}-7F0+4F1-2f_{m2}+f_{S7}$ |
| 147 | 35.0051 ± 0.0005 | 0.175 | 6.4 | $4F1+4f_{S5}-6F0-f_{m1}-f_{S6}-f_{m2}$ |
| 148 | 16.0114 ± 0.0002 | 0.363 | 17.7 | $4F1+4f_{S5}-7F0-f_{m1}-f_{S6}-f_{m2}$ |
| 149 | 5.3288 ± 0.0005 | 0.159 | 4.5 | $7F0+f_{S6}+f_{m1}+2f_{m2}-3F1-f_{S7}-3f_{S5}$ |
| 150 | 21.4558 ± 0.0003 | 0.229 | 7.0 | $7f_{S5}-6F0-3f_{m1}-f_{S7}+F1$ |
| 151 | 2.9824 ± 0.0004 | 0.200 | 7.9 | $8F0-4F1-4f_{S5}+f_{m1}+f_{S6}+f_{m2}$ |

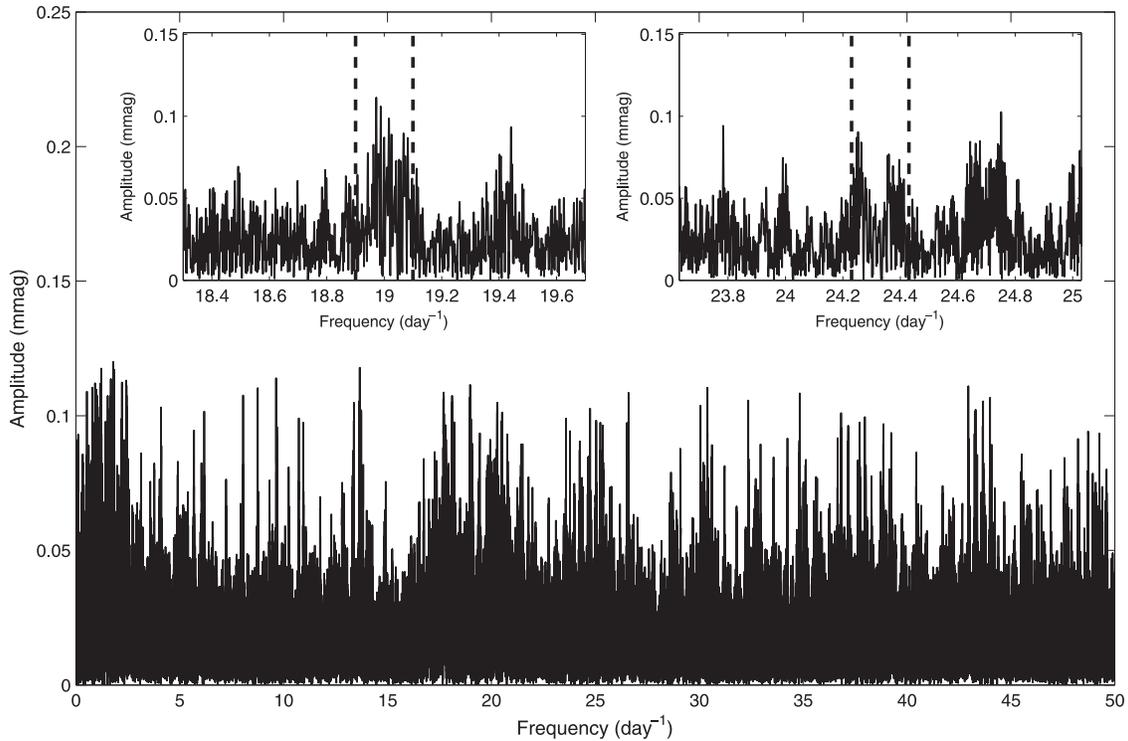

**Figure 2.** Residuals left behind after extraction of 151 significant frequencies, to amplitude limits 0.12 mmag. No peak is statistically significant in the residuals. Zooms into the region around the main frequencies F0 (left inset) and F1 (right inset) are marked with vertical dashed lines, showing the multiplet structures.

independent pulsation modes. Moreover, a quintuplet, which is assumed to be stellar rotationally split, is detected in the frequency spectra, and its separation is nearly equal, with a mean separation of 0.218 day$^{-1}$ (Murphy et al. 2013). However, it is different from that in KIC 10284901, as the quintuplet in this work includes two exact interval, i.e., $f_{m1} = 0.4407$ day$^{-1}$ and $f_{m2} = 0.8125$ day$^{-1}$. From this perspective, these two quintuplets may have different origins.

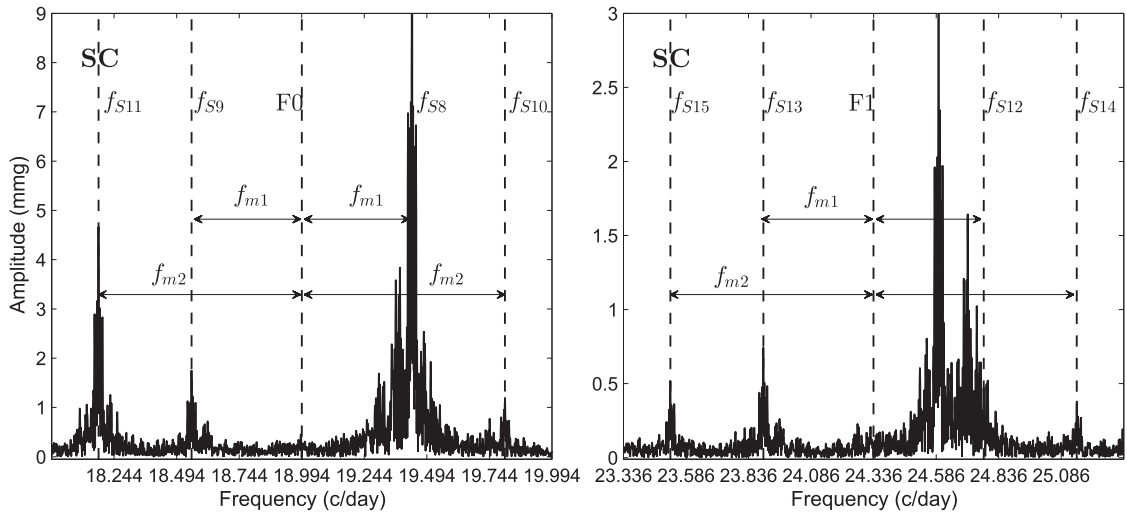

**Figure 3.** Amplitude spectra after subtracting the main frequencies. The vertical dashed lines indicate the locations of the main frequencies F0 (left panel), F1 (right panel), and their multiplets. The two panels clearly show four pairs of side peaks ( 1: $f_{S8}, f_{S9}$ 2: $f_{S10}$ and $f_{S11}$ centered on F0; 3: $f_{S12}, f_{S13}$; 4: $f_{S14}$ and $f_{S15}$ centered on F1) around the main frequency F0 and F1 in SC spectrum. The intervals from the side peaks to the center are marked with $f_{m1}$ and $f_{m2}$, respectively.

Yang et al. (2018b) reported that a pair of equidistant side peaks around the main frequency were detected in the frequency spectra in a HADS star KIC 5950759 and they were caused by the the amplitude modulation to its main pulsation with a stellar rotation of 0.3193 day$^{-1}$ ($v \sin i \approx 33$ km s$^{-1}$). In the case of KIC 10284901, the possibility of amplitude modulation with rotation was considered to explain the multiplet structures; however, it is hard to imagine how the stellar rotation, and the low rotation velocity commonly found in HADS stars, could produce two different modulation frequencies, i.e., $f_{m1} = 0.4407$ day$^{-1}$ and $f_{m2} = 0.8125$ day$^{-1}$. As a result, there is not enough evidence to support the hypothesis that these two modulations are from the stellar rotation.

### 4.1.3. Blazhko-like Effect?

In Blazhko RR Lyrae stars, the equidistant triplets are often shown in the frequency spectra and the interval of the triplets is usually equal to the modulation frequency (Blažko 1907; Kolenberg et al. 2006; Soszyński et al. 2016). The modulation frequency can also be detected directly in the frequency spectra. Hurta et al. (2008) and Kolenberg et al. (2011) also found equidistant quintuplets in the spectrum of RR Lyrae star.

Jurcsik et al. (2014) presented a detailed analysis of four double-mode RR Lyrae stars showing the Blazhko effect based on new time-series photometry of the globular cluster M3. These four double-mode stars show large-amplitude Blazhko modulations of both radial modes and rapid phase change connected to the amplitude minimum of the respective mode. One of them, V13, shows a anti-correlation in both the amplitude- and phase- modulations of the modes. In KIC 10284901, the temporal behaviors of the amplitudes and phases of the two radial pulsation modes were investigated based on the observational and synthetic data. We took the 151 frequency fit to the full data and then computed a synthetic light curve to the full data set using the parameters from this fit, but leaving out F0, F1, and the T1 and T2 groups. Then the difference of the original data and this fit was fitted by F0 and F1 only in a short pieces (the time interval of 0.5 day was chosen to obtain a much better time resolution). The amplitude and phase variations of the modes are shown in Figure 4. Contrary to the anti-correlation shown in V13 (see Figure 2 in Jurcsik et al. 2014), the amplitudes of the two modes in KIC 10284901 seem to show obviously cyclic change and synchronous behavior. Fourier analysis of the amplitude variations of these two modes reveals two significant frequencies of 0.44071(1) day$^{-1}$ and 0.81243(4) day$^{-1}$, which are equal to $f_{m1}$ and $f_{m2}$. Like its amplitude, the phase of mode F0 also shows obviously cyclic change, but the mode F1 does not possess similar variation.

Smolec et al. (2015) gave an analysis of Blazhko-type modulation in double-mode RR Lyrae stars in the Optical Gravitational Lensing Experiment photometry of the Galactic bulge, and found the amplitudes and phase of the radial modes varied irregularly on a long timescale of a few hundred or thousand days; the same is true for the short-term modulation. In Figure 4, the variations of the amplitude of both the radial modes, which are commonly in the Blazhko RR Lyrae stars (e.g., CoRoT 105288363: Guggenberger et al. 2011 and V445 Lyr: Guggenberger et al. 2012), are obvious and regular. From this perspective, the modulation detected in KIC 10284901 seems to differ from that in Blazhko RR Lyrae stars. However, in Figure 2 the multiplet structures in the main pulsations in the residual spectrum display a similarity to those in RR Lyrae star CoRoT 105288363 (Guggenberger et al. 2011), and the amplitudes of the modes of KIC 10284901 also show strong variations, as shown in CoRoT 105288363.

Benkó et al. (2014) reported that three Blazhko RR Lyrae stars in the *Kepler* field (i.e., V355 Lyr, V 366 Lyr, V450 Lyr) show two modulation periods in their light curves, and the ratio between the primary and secondary modulation periods is nearly 1:2. In KIC 10284901, the SC light curve exhibits an obvious modulation feature in its amplitude, as commonly shown in the Blazhko RR Lyrae stars (e.g., Benkó et al. 2014). The multiplet structures found in the SC spectrum, as shown in Figure 3, are similar to that in the Blazhko RR Lyrae stars. Moreover, the modulation frequencies ($f_{m1}$ and $f_{m2}$), which were detected directly in SC data, have a ratio of nearly 1:2 just as shown in the above Blazhko RR Lyrae stars. These features

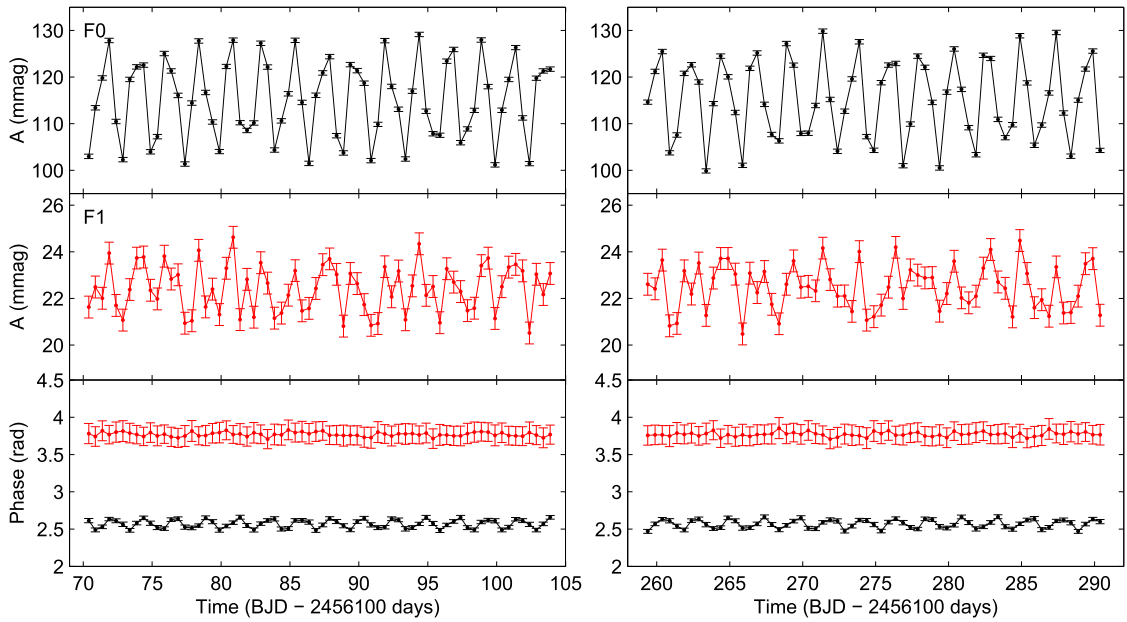

**Figure 4.** Temporal behavior of the amplitude and phase variations of the two radial pulsation modes in KIC 10284901. The top and middle panels show the amplitude variations of F0 (in black) and F1 (in red), respectively. The bottom panels show the phase variations of F0 (in black) and F1 (in red), respectively. For clarity, we only show the data from BJD 2456170–2456205 days (left panels), and BJD 2456355–2456400 days (right panels). Each point contains a non-overlap 0.5 day subset derived from the difference of the original light curve and the synthetic data.

shown in KIC 10284901 are similar to the frequency patterns in the Blazhko RR Lyrae stars.

Thus, for the target KIC 10284901, the features, including the multiplet structures in the frequency spectrum, the ratio of $f_{m1}$ and $f_{m2}$ (nearly 1:2), the obviously cyclic variations of the amplitude of the two dominant pulsation modes, are similar to the Blazhko effect in RR Lyrae stars. It seems to imply the modulation is related to the Blazhko effect. As the phase variation of the mode F1 seems to not be obvious, note that although the modulation detected in KIC 10284901 shares some similarities with the Blazhko effect, it could also belong to a unique effect of HADS stars, which would require further investigation. More HADS stars with this modulation effect are needed to solve this mystery, especially from space missions.

### 4.2. Location in the H–R Diagram

To investigate the evolutionary stage of KIC 10284901, we obtained $M_V = 2.84(\pm0.17)$ mag for this star using the period–luminosity relationship $M_V = -1.83(\pm0.08) - 3.65(\pm0.07) \log P_F$ ($P_F$ is the period of the fundamental mode) provided by Poretti et al. (2008) and its $\log P_F = -1.279$. Thus, in the H–R diagram, KIC 10284901 is located on the bottom of the HADS instability strip and likely situated in the main sequence, under the constraints of the above $M_V = 2.84(\pm0.17)$ mag and $T_{\rm eff} = 7710(\pm180)$ K derived from GTC spectra in the RATS-Kepler by Ramsay et al. (2014). Note that a precise parallax (0.328 ± 0.023 mas) and interstellar extinction value ($A_G = 0.21$ mag) for this star are available in the Gaia second Data Release (DR2)[5] (Gaia Collaboration et al. 2018), and the resulting distance ($d_{\rm parallax} = 3064$ (±215) pc) is very consistent with the results ($d = 3002$ (±235) pc) derived from the period–luminosity relationship by Poretti et al. (2008).

---

[5] ESA Gaia Archive: http://gea.esac.esa.int/archive/.

### 5. Summary

We analyzed the pulsations of Kepler target KIC 10284901, and extracted 151 significant frequencies from the SC data. Among them, seven independent frequencies were found in the SC spectrum. The period ratio (=0.7805) of the first overtone (F1) and fundamental mode (F0) suggests that this star is a double-mode HADS star. The derived absolute visual magnitude $M_V = 2.84 \pm 0.17$ mag, as well as the effective temperature $T_{\rm eff} = 7710 \pm 180$ K obtained from GTC spectra, indicate that KIC 10284901 lies in the bottom of the HADS instability region and is likely a main-sequence star. With the precise parallax provided by Gaia, the distance of KIC 10284901 is derived as 3064 (±215) pc.

KIC 10284901 is the first double-mode HADS star in which the quintuplet structures around the main pulsation modes were detected in the frequency spectra. The quintuplet structures are caused by two modulation frequencies, $f_{m1} = 0.4407$ day$^{-1}$ and $f_{m2} = 0.8125$ day$^{-1}$. The temporal behavior of the amplitude of the two dominant modes reveals that the amplitudes vary with the same trend, and the phase of F0 also shows obviously cyclic change. The features of the frequency patterns, the ratio ($f_{m1}/f_{m2} \approx 1:2$) of the two modulation frequencies, and the obviously cyclic variations of amplitude of the two dominant pulsation modes, suggest that the modulation in this star might be related to the Blazhko effect. Nonetheless, the possibility that this modulation just belongs to HADS stars cannot be ruled out completely, which is worth further study. More HADS stars with this modulation are needed to verify its nature and investigate the relationship between HADS stars and RR Lyrae stars, and further investigations of this modulation could provide a new perspective on the classical instability strip in the H–R diagram.

The authors thank the referee for the very helpful comments and the editor for the careful revision of the manuscript. This research is supported by the program of the National Natural

Science Foundation of China (grant No.11873081). We thank the *Kepler* science team for providing such excellent data.

## References

Andrae, R., Fouesneau, M., Creevey, O., et al. 2018, A&A, 616, A8
Balona, L. A. 2014, MNRAS, 437, 1476
Balona, L. A., & Dziembowski, W. A. 2011, MNRAS, 417, 591
Balona, L. A., Lenz, P., Antoci, V., et al. 2012, MNRAS, 419, 3028
Bedding, T. R., Mosser, B., Huber, D., et al. 2011, Natur, 471, 608
Benkő, J. M., Plachy, E., Szabó, R., et al. 2014, ApJS, 213, 31
Blažko, S. 1907, AN, 175, 325
Borucki, W. J., Koch, D., Basri, G., et al. 2010, Sci, 327, 977
Bowman, D. M., & Kurtz, D. W. 2014, MNRAS, 444, 1909
Bowman, D. M., Kurtz, D. W., Breger, M., et al. 2016, MNRAS, 460, 1970
Breger, M. 2000, in ASP Conf. Ser. 210, Delta Sciti and Related Stars (San Francisco, CA: ASP), 3
Breger, M., Balona, L., Lenz, P., et al. 2011, MNRAS, 414, 1721
Breger, M., & Kolenberg, K. 2006, A&A, 460, 167
Breger, M., Stich, J., Garrido, R., et al. 1993, A&A, 271, 482
Brown, T. M., Latham, D. W., Everett, M. E., et al. 2011, AJ, 142, 112
Gaia Collaboration, Brown, A. G. A., Vallenari, A., et al. 2018, A&A, 616, A1
Giammichele, N., Charpinet, S., Fontaine, G., et al. 2018, Natur, 554, 73
Gilliland, R. L., Brown, T. M., Christensen-Dalsgaard, J., et al. 2010, PASP, 122, 131
Greiss, S., Steeghs, D., Gänsicke, B. T., et al. 2012, AJ, 144, 24
Guggenberger, E., Kolenberg, K., Chapellier, E., et al. 2011, MNRAS, 415, 1577
Guggenberger, E., Kolenberg, K., Nemec, J. M., et al. 2012, MNRAS, 424, 649
Hurta, Z., Jurcsik, J., Szeidl, B., & Sódor, Á 2008, AJ, 135, 957
Jurcsik, J., Smitola, P., Hajdu, G., & Nuspl, J. 2014, ApJL, 797, L3
Kjeldsen, H., Christensen-Dalsgaard, J., Handberg, R., et al. 2010, AN, 331, 966
Koch, D. G., Borucki, W. J., Basri, G., et al. 2010, ApJL, 713, L79
Kolenberg, K., Bryson, S., Szabó, R., et al. 2011, MNRAS, 411, 878
Kolenberg, K., Smith, H. A., Gazeas, K. D., et al. 2006, A&A, 459, 577
Lenz, P., & Breger, M. 2005, CoAst, 146, 53
McNamara, D. H. 2000, in ASP Conf. Ser. 210, Delta Scuti and Related Stars, ed. M. Breger & M. H. Montgomery (San Francisco, CA: ASP), 373
Montgomery, M. H., & Odonoghue, D. 1999, DSSN, 13, 28
Murphy, S. J., Pigulski, A., Kurtz, D. W., et al. 2013, MNRAS, 432, 2284
Pamyatnykh, A. A. 2000, in ASP Conf. Ser. 210, Delta Scuti and Related Stars, ed. M. Breger & M. Montgomery (San Francisco, CA: ASP), 215
Petersen, J. O., & Christensen-Dalsgaard, J. 1996, A&A, 312, 463
Poretti, E., Clementini, G., Held, E., et al. 2008, ApJ, 685, 947
Ramsay, G., Brooks, A., Hakala, P., et al. 2014, MNRAS, 437, 132
Smolec, R., Soszyński, I., Udalski, A., et al. 2015, MNRAS, 447, 3756
Soszyński, I., Smolec, R., Dziembowski, W. A., et al. 2016, MNRAS, 463, 1332
Stellingwerf, R. F. 1979, ApJ, 227, 935
Yang, T. Z., Esamdin, A., Fu, J. N., et al. 2018a, RAA, 18, 2
Yang, T. Z., Esamdin, A., Song, F. F., et al. 2018b, ApJ, 863, 195